\documentclass[11pt,a4paper]{article}
\usepackage{jheppub}
\usepackage[dvipsnames]{xcolor}

\newcommand{\eq}{\begin{equation}}
\newcommand{\eqe}{\end{equation}}

\newcommand*\dd{\mathop{}\!\mathrm{d}}

\newcommand{\eqa}{\begin{eqnarray}}
\newcommand{\eqae}{\end{eqnarray}}

\def\tr{\text{tr}} 

\def\R{\mathbb{R}}

\def\<{\langle}
\def\>{\rangle}
\def\+{\dagger}

\def\Oo{{\cal O}}

\begin{document}

\title{Operator thermalisation in $d>2$: Huygens or resurgence}

\author[a]{Julius Engels\"oy,}
\emailAdd{julius.engelsoy@fysik.su.se}

\author[a]{Jorge Larana-Aragon,}
\emailAdd{jorge.laranaaragon@fysik.su.se}

\author[a]{Bo Sundborg,}
\emailAdd{bo@fysik.su.se}
\affiliation[a]{The Oskar Klein Centre for Cosmoparticle Physics \& Department of Physics, Stockholm University, \\
AlbaNova, 106 91 Stockholm, Sweden.\\}

\author[b]{Nico Wintergerst}
\emailAdd{nico.wintergerst@nbi.ku.dk}
\affiliation[b]{The Niels Bohr Institute, University of Copenhagen, \\
Blegdamsvej 17, 2100 Copenhagen Ø, Denmark\\}

\abstract{
Correlation functions of most composite operators decay exponentially with time at non-zero temperature, even in free field theories. This insight was recently codified in an OTH (operator thermalisation hypothesis). We reconsider an early example, with large $N$ free fields subjected to a singlet constraint. This study in dimensions $d>2$ motivates technical modifications of the original OTH to allow for generalised free fields. Furthermore, Huygens' principle, valid for wave equations only in even dimensions, leads to differences in thermalisation. It works straightforwardly when Huygens' principle applies, but 
thermalisation is more elusive if it does not apply. Instead, in odd dimensions we find a link to resurgence theory by noting that  exponential relaxation is analogous to non-perturbative corrections to an asymptotic perturbation expansion. Without applying the power of resurgence technology we still find support for thermalisation in odd dimensions, although these arguments are incomplete.
}
\maketitle

\section{Introduction}\label{sec:intro}

Free field theories are the simplest and most prominent examples of (super-)integrable quantum field theories (QFTs), rendered exactly solvable by the existence of an infinite set of conserved charges. A direct consequence of the presence of such charges is a severely constrained time evolution even in thermal backgrounds. In particular, simple operators in free QFTs fail to satisfy the requirements of the eigenstate thermalisation hypothesis \cite{Deutsch1991,Srednicki1994} and their late time behaviour is therefore unlikely to approach ensemble averages, tantamount to the absence of \emph{thermalisation}.

Nonetheless, it is known that nontrivial interference effects can effectively mimic equilibration. For example, after quantum quenches \cite{CalabreseCardy2006,CalabreseCardy2007,DasGalanteMyers2016,BanerjeeEngelsoyLarana-ArSundborgThorlaciuWinterger2019}, correlation functions in free QFTs approach those of a generalised Gibbs ensemble \cite{RigolDunjkoYurovskyOlshanii2007,CalabreseCardy2007,Cardy2016,DymarskyPavlenko2019}, characterised by chemical potentials for all conserved charges which in free QFTs is equivalent to a momentum-dependent temperature. Similarly, nontrivial time dependence arises when considering \emph{composite operators}. Such operators can in fact interact with the thermal bath and as such exhibit a range of phenomena that are usually attributed to their interacting counterparts. For instance, their correlation functions can exhibit exponential decay at late times \cite{AmadoSundborgThorlaciusWintergerst2018} and their spectral densities have support in the deeply off-shell regime \cite{AmadoSundborgThorlaciusWintergerst2018,BanerjeePapadodimRajuSamantrayShrivasta2020}, reminiscent of collision-less Landau damping \cite{SonStarinets2002}.

Clearly, the composite nature of an operator is a necessary condition for its effective thermalisation, since only then does it couple to a thermal bath, indicated by a temperature dependence of its response functions. On the other hand, to which extent it is also a sufficient condition is less understood. In recent work \cite{Sabella-GarnierSchalmVakhtelZaanen2019,BukvaSabella-GarnierSchalm2019}, a simple criterion has been formulated that guarantees the \emph{absence of thermalisation} of a given operator, characterised by a lack of exponentially decaying contributions to its linear response function. In addition, it was conjectured that a converse statement can be made and any operator that fails this non-thermalisation condition in fact thermalises. This conjecture was introduced as the Operator Thermalisation Hypothesis (OTH).

This note aims to shed light on several remaining puzzles. First, in singlet models \cite{AmadoSundborgThorlaciusWintergerst2018}, the  calculated correlation functions were observed to display exponential decay in even dimensions $d > 2$. This decay is directly related to the thermalisation later extracted in \cite{Sabella-GarnierSchalmVakhtelZaanen2019} by arguments which however do not resolve a difference between even and odd dimensions. Since the odd-dimensional singlet model correlation functions do not decay exponentially, the results appear to be in tension with each other. There is something to learn about thermalisation or singlet models, in fact both, from a closer study. The requisite developments of concepts indeed leads to a more precise formulation of the OTH. Second, the singlet model study demonstrated how phases below or above a critical temperature exhibit different relaxation properties, most plainly for auto-correlators. Since response functions diagnose thermalisation, previous singlet model studies should be extended with results on response functions in different phases.

Thus, we will study OTH in a particular class of free field theories, namely those with a large-$N$ singlet constraint. These theories have received widespread attention in the context of gauge/gravity duality as the holographic duals of gravitational theories with an infinite tower of massless fields of higher spin. They exhibit an interesting thermal structure on compact spaces with a large $N$ confinement/deconfinement phase transition. In ordinary AdS/CFT, this transition is also present and can be mapped to the Hawking--Page transition from thermal AdS to the large AdS black hole in the bulk. In the deconfined phase, thermalisation in holographic gauge theories is in direct correspondence with black hole formation and equilibration in the bulk. Understanding thermal properties of free singlet models thus provides insight on putative black holes in higher spin gravity. 
More generally, however, they allow one to disentangle generic properties of composite operators from those particular to strong coupling, thereby teaching valuable lessons on the inner workings of gauge/gravity duality.

In the low temperature phase we observe the absence of thermalisation to leading order in $1/N$, in complete accordance with the OTH. Below the phase transition, a composite operator $\tr(\Phi(x)\Phi(x))$, built of $N$ adjoint scalars, plays the role of a generalised free field with interaction strength of order $1/N$. To leading order, it obeys our generalisation of the non-thermalisation condition, confirmed by the absence of temperature dependent contributions to its response functions and in particular the lack of exponential damping. This is generic to all QFTs that admit a description in terms of generalised free fields and the thermal version of this concept will be presented below. At high temperatures, the time dependence becomes significantly richer. Response functions become temperature dependent and are characterised by non-analyticities off the real axis in the complex frequency plane. They describe a damped response to sources, with a power law tail and sub-leading exponentially decaying contributions. The latter contribution is the exponential damping predicted by the OTH. 
The presence of the power law tail implies that information about the source is retained to a larger degree than in standard thermalisation, although parts are effectively lost in exponentially damped terms. 

The general lessons from our study concern details of the formulation of OTH, and the difference between even and odd dimensions. Indeed, it is well known that the interior of the light cone plays a fundamentally different role in wave propagation in even and odd dimensions (cf Huygen's principle and Hadamard's problem \cite{Gunther1991}). By explicitly focusing on evaluating correlators close to the light cone we reduce the difference between odd and even dimensions, and identify the damped quantities that continue analytically  between different dimensions, to put $d>2$ OTH on a firmer footing. This light cone limit notwithstanding, crucial differences between even and odd dimensions remain. While OTH can be confirmed straightforwardly in even dimensions, we observe that subtleties involved in isolating exponentially decaying terms in the response functions become critical in odd dimensions. We describe the difficulties and find some support for thermalisation, but also indications that the resolution requires more powerful tools from the theory of resurgence \cite{Ecalle1981,Marino2014,AnicetoBasarSchiappa2019,Dorigoni2019}.  That cautionary observation aside, our scrutiny of OTH permits us to give a more precise formulation of both the hypothesis and the converse non-thermalisation condition in all $d>2$.

Our paper is organised as follows. In section \ref{sec:oth}, we introduce the concept of operator thermalisation, non-thermalisation and the role stable thermal quasi-particles and generalised free fields. Before going into basics of singlet models in \ref{sec:singlet models} we also introduce the potential relation of thermalisation to resurgence. In section \ref{sec:lowT}, we then deduce and discuss absence of exponential relaxation in singlet model response functions in the low temperature phase. The high temperature phase, which displays relaxation in even dimensions and appears to allow for it in odd dimensions, is analysed in section \ref{sec:highT} and a discussion in section \ref{sec:discussion} leads up to our conclusions \ref{sec:conclusions}.

\section{Preliminaries}\label{sec:prel}

\subsection{Operator thermalisation}\label{sec:oth}

Sabella-Garnier et al formulated the operator thermalisation hypothesis in \cite{Sabella-GarnierSchalmVakhtelZaanen2019} and considered the thermalisation properties of operator correlation functions in a fixed background rather than operator expectation values in the presence of of assumptions on the energy spectrum,
as done by the eigenstate thermalisation hypothesis, ETH  \cite{Deutsch1991,Srednicki1994}. They discuss thermalisation in terms of an exponentially fast return to equilibrium of operator expectation values in response to a perturbation by the operator in question. More precisely, in \cite{Sabella-GarnierSchalmVakhtelZaanen2019} the retarded Green's function of the operator in question is taken to define \emph{thermalisation of a perturbation}, when it decays exponentially, in line with the retarded Green's function encoding the linearised response of the operator $\mathcal{O}$ induced by a  perturbation by the same operator $\mathcal{O}$. 
The requirement of exponential decay for a perturbation to thermalise corresponds to the intuition that a thermalising perturbation is ``forgotten'' by the system at late times. The latter means that exponential precision would be required in order to fully reconstruct the source from the response of the medium.

A motivation behind the operator thermalisation hypothesis, and one of its strengths, is that it can be used to study surprising similarities between ordinary interacting systems and free or integrable systems \cite{AmadoSundborgThorlaciusWintergerst2018,BanerjeePapadodimRajuSamantrayShrivasta2020,BanerjeeEngelsoyLarana-ArSundborgThorlaciuWinterger2019,Sabella-GarnierSchalmVakhtelZaanen2019}. While pure exponential decay occurs in free systems in contrast to naive expectations, it is generally masked by leading power law decay for $d>2$, as well as multiplied by inverse powers of time. We will provide such examples below. In reviewing the operator thermalisation hypothesis, we will therefore introduce new terminology which precisely captures these features. In effect, we demonstrate an operator non-thermalisation condition which excludes this kind of \emph{partial} thermalisation, and state a converse \emph{partial} operator thermalisation hypothesis. Our arguments are essentially copied from  \cite{Sabella-GarnierSchalmVakhtelZaanen2019}, and the ``partial'' qualifier only indicates a slight shift of definitions. The new definitions are important for consistency with the examples we discuss, but the idea is approximately the same.

\subsubsection{Partial operator thermalisation}

We define \emph{partial thermalisation} of an operator $\mathcal{O}$ to mean that: The retarded Green's function of $\mathcal{O}$ contains terms with exponentially damped factors at late times. 
This definition allows for leading power-law decay, and exponential terms which are only sub-leading\footnote{
Identifying sub-leading exponentials is subtle and requires special attention (provided in \ref{sec:resurgence}).}. In such cases, time evolution still ``forgets'' part of the initial perturbation, but not all of it. Clearly, partial thermalisation includes the thermalisation notion discussed in \cite{Sabella-GarnierSchalmVakhtelZaanen2019} and the more conventional notion of approach to a thermal ensemble, but it is a broader concept\footnote{In private discussions, we have found that the authors of \cite{Sabella-GarnierSchalmVakhtelZaanen2019} are aware of the need for some refinements.}. 

Crucially, partial operator thermalisation captures the observation that conservation laws prevent some operators in free or integrable theories to thermalise, but that almost all other operators thermalise partially.   A special class of \emph{non-thermalising operators} was characterised by Sabella-Garnier et al \cite{Sabella-GarnierSchalmVakhtelZaanen2019}.
 We will see that these operators \emph{do not even thermalise partially}. In essence, these non-thermalising operators are generalisations of free fields which satisfy a sharp dispersion relation relating energy to momentum.
Formally, the conditions on the operators are given by the mathematical descriptions below. Physically, they correspond to stable thermal quasi-particle fields having clear-cut dispersion relations, which are permitted to differ from those of free relativistic particles. One may invoke the Narnhofer-Requardt-Thirring theorem \cite{NarnhoferRequardtThirring1983} to argue that they describe a sector of the thermal system which is  completely free from interactions, except for modified dispersion relations. The theorem permits other sectors, but they are completely decoupled from the quasi-particles.

We interpret the operator thermalisation hypothesis proposed in \cite{Sabella-GarnierSchalmVakhtelZaanen2019} to state that any other local operator, not representing a stable quasi-particle field, will thermalise. This is the converse of the above non-thermalisation condition. For it to hold, the notion of thermalisation has to be weakened to partial thermalisation. Thus, we propose a more precise \emph{partial operator thermalisation hypothesis}: Any local operator not representing what we call a thermal generalised free field\footnote{Generalised free fields were originally introduced in \cite{Greenberg1961}, and also play a natural role in AdS/CFT as boundary duals of free fields in AdS.}, or a generalised quasi-particle field, thermalises partially. Note that we still have not proven this hypothesis, though we find it reasonable. All the plausibility arguments in \cite{Sabella-GarnierSchalmVakhtelZaanen2019} still apply. 

\subsubsection{Non-thermalisation and the thermalisation hypothesis} \label{sec:non-th}

We now proceed to essentially repeat the arguments of \cite{Sabella-GarnierSchalmVakhtelZaanen2019}, expressed in our terminology.

Consider a stable thermal quasi-particle operator, which we denote ${\cal Q}(t,x)$ to distinguish it from more general local operators ${\cal O}(t,x)$. By definition it has a definite dispersion relation. 
In finite volume and in a basis which simultaneously diagonalises energy and momentum, this means that the transitions ${\cal Q}$ can mediate between momentum states determine the simultaneous transitions between energy eigenvalues. We do not need to know if there is a single functional relation between momentum and energy for the operator  ${\cal Q}$ or if there are several branches of solutions to the dispersion relations coupling to ${\cal Q}$. To reproduce branch cuts which can be found, for example, in singlet models, it will turn out to be important to allow for a growth of the number of solutions to dispersion relations with volume.

The retarded thermal Green's function is
\begin{equation}
G_{R}(t,x)=-i \Theta(t) \left\langle  \left[ {\cal Q}(t,x), {\cal Q}(0,0)\right] \right\rangle_{\beta}
\end{equation}
which can be expanded in a sum of expectation values
\begin{equation}
G_{R}(t,x)= -i \frac{\Theta(t)}{Z(\beta)} \sum_{n} e^{{-{\beta E_{n}}}} 
\left\{ \left\langle n \right| {\cal Q}(t,x) {\cal Q}(0,0) \left| n \right\rangle 
 - (-1)^{2s}\left\langle n \right| {\cal Q}(0,0) {\cal Q}(t,x) \left| n \right\rangle \right\},
\end{equation}
where $s$ is the spin of the operator ${\cal Q}$. Making use of translations
\begin{equation}
{\cal Q}(t,x) = e^{iHt}e^{-iP\cdot x} {\cal Q}(0,0) e^{-iHt}e^{iP\cdot x}
\end{equation}
and inserting a complete set of states
\begin{multline}
G_{R}(t,x)= -i \frac{\Theta(t)}{Z(\beta)} \\\times \sum_{m,n} e^{{-{\beta E_{n}}}} 
\left\{ 
 e^{-i(E_{m} - E_{n})t + i (p_{m} - p_{n})\cdot x} 
 - (-1)^{2s} e^{i(E_{m} - E_{n})t - i (p_{m} - p_{n})\cdot x}
 \right\} 
 \left| \left\langle m \right| {\cal Q} \left| n \right\rangle \right|^{2}.
\end{multline}
In Fourier space 
\begin{multline}
G_{R}(\omega,k) = - \frac{i}{Z(\beta)} \\ \times \sum_{m,n} e^{{-{\beta E_{n}}}}
\left\{
\frac{ \delta\left( k -(p_{n}-p_{m}) \right) }{ \omega + (E_{m} - E_{n}) + i\epsilon }
- (-1)^{2s} \frac{ \delta\left( k -(p_{m}-p_{n}) \right) }{ \omega + (E_{n} - E_{m}) + i\epsilon }
\right\} 
\left| \left\langle m \right| {\cal Q} \left| n \right\rangle \right|^{2},
\end{multline}
where $\epsilon > 0 $ is infinitesimal.

Now, the special properties of the quasi-particle operator $ {\cal Q} $ lead to a proof of non-thermalisation. Denoting by $M$ the number of 
different branches of solutions 
\begin{equation}
\omega = \Omega_{j}^{{\cal Q}}(k) \label{eq:dispersion}
\end{equation}
labeled $j =1,\ldots,M$ to the dispersion relations for $ {\cal Q} $ and defining the residue functions
\begin{equation}
H_{j}^{{\cal Q}}(\beta,k) = - \frac{i}{Z(\beta)} \sum_{E_{m}-E_{n} = \Omega_{j}^{{\cal Q}}(p_{m}-p_{n})} e^{{-{\beta E_{n}}}}
\delta\left( k -(p_{n}-p_{m}) \right) 
\left| \left\langle m \right| {\cal Q} \left| n \right\rangle \right|^{2},
\end{equation}
we find
\begin{equation} \label{eq:dispersion response}
G_{R}(\omega,k) = \sum_{j=1}^{M} 
\left\{
\frac{ H_{j}^{{\cal Q}}(\beta,k) }{ \omega - \Omega_{j}^{{\cal Q}}(k) + i\epsilon }
- (-1)^{2s} \frac{ H_{j}^{{\cal Q}}(\beta,-k) }{ \omega + \Omega_{j}^{{\cal Q}}(-k) + i\epsilon }
\right\} .
\end{equation}
The frequencies and wave numbers above are related to the matrix elements $\left\langle m \right| {\cal Q} \left| n \right\rangle$ by
\begin{equation}
E_{m}-E_{n}=\Omega_{j}^{{\cal Q}}(p_{m}-p_{n}), \label{eq:matrix elements}
\end{equation}
signifying that all contributions from the operator ${\cal Q}$ are due to transitions between states whose energies and momenta differ by amounts related by the allowed dispersion relations in eq.\ \eqref{eq:dispersion}. 
For a more detailed analysis of thermodynamic and large $N$ limits, it may become useful to allow an effective temperature dependence in the dispersion relations contributing to eq.\ \eqref{eq:dispersion response}. 
Noting that $ \Omega_{j}^{{\cal Q}}(k) $ has to be real by definition, the retarded Green's function only has singularities on the real axis, which is tantamount to non-thermalisation of the operator $ {\cal Q} $. For finite $M$, the singularities are manifestly poles. If $M$ grows without bound in the thermodynamic or large $N$ limit,  branch cuts may also arise, but they will be on the real axis. There will still not even be partial thermalisation, since only singularities off the real axis can produce exponentially decaying terms.

The converse of the original non-thermalisation result would be that only stable quasi-particle operators are non-thermalising. Allowing for partial thermalisation, which includes power law fall-offs related to branch cuts on the real axis, it seems judicious to consider branch cuts also in the non-thermalisation results. Thus we are led to allow unbounded $M$. This generalisation replaces quasi-particles with generalised quasi-particles or thermal generalised free fields.

The OTH in our version becomes: All local operators which are not generalised quasi-particles thermalise partially. The original plausibility arguments of \cite{Sabella-GarnierSchalmVakhtelZaanen2019} remain, and this adjusted version survives all tests we have considered.

\subsubsection{Thermalisation and resurgence}\label{sec:resurgence}

Below we will introduce examples of retarded Green's functions with asymptotic late time expansions containing both inverse powers and damped exponentials of time. In free systems, they force us to consider the partial, and more general, version of operator thermalisation, which allows for the possibility that exponential damping terms are sub-leading. Unfortunately, the price for the generalisation is another level of mathematical sophistication. It is required for a physical reason: Only under very special circumstances, e.g. when an asymptotic series of inverse powers terminates, is it possible to operationally separate sub-leading exponentials from more important inverse powers. Only under these special circumstances can we have a chance to resolve and observe the damped exponentials, even in principle. 

This discussion is parallel to the potentially more familiar discussion about prescription dependence of non-perturbative terms in quantum mechanics and in quantum field theory. There, one encounters non-perturbative exponentials $e^{-1/g^{2}}$ complementing power series in a coupling $g$. Substituting 
\begin{eqnarray*}
g^{2} &\to& \beta/t  \\
e^{-1/g^{2}} &\to& e^{-t/\beta}\, ,
\end{eqnarray*}
where $t$ is time and $\beta$ is inverse temperature, we are alerted to the possibility that thermalisation, signalled by exponential damping at late times, can be analogous to non-perturbative effects. The analogy indeed holds for standard Green's functions: Their late time expansion in inverse powers of $t/\beta$ is typically asymptotic rather than convergent, and exponential terms can sometimes be extracted from integral representations of the Green's functions. Cases with terminating or at least a convergent (inverse) power series would be useful in practice, and would allow unambiguous identification of exponentials, but are exceptional.

The beautiful idea that there is a relation between the form of non-perturbative terms and the divergence of asymptotic series \cite{Lipatov1977} can be systematised in non-perturbative techniques like Borel resummation, but does not always yield a unique answer for the series. To be clear, for thermal Green's functions in free field theory, the integral representations are unambiguous. A series representation does not improve the already complete encoding of a response function. However, a well-defined representation of the result of the integral in a double series expansion with inverse powers and exponentials as above, a trans-series in the framework of resurgence theory \cite{Marino2014,AnicetoBasarSchiappa2019,Dorigoni2019}, would lend itself nicely to an extended definition of partial thermalisation. The response function would be said to thermalise partially if the series contained exponentials\footnote{Such a subtle definition may seem outlandish, but is apparently useful in the mathematical description of related physics, namely classical Landau damping \cite{MouhotVillani2011}.}.

\subsection{Thermal singlet models}\label{sec:singlet models}

In order to distinguish low and high temperatures, we consider free field theories on $\R \times S^{d-1}$ leading to a characteristic temperature scaling as $1/R$, the inverse of the radius $R$ of the sphere $S^{d-1}$. To make the distinction sharper we consider a large number $N$ of fields. A large $N$ will then allow for qualitatively different limits for physics below and above the characteristic temperature. We consider a scalar field transforming in a representation, usually fundamental or adjoint, of some large $N$ symmetry group, for example U$(N)$ or O$(N)$. Projection onto the singlet sector is achieved by weakly gauging the symmetry, i.e.\ introducing a gauge field $A_\mu$ in the limit of vanishing gauge coupling, where only the zero mode $\alpha \sim \int_{S^{d-1}} A_0$ that imposes the Gauss' law constraint remains. We will focus attention on correlation functions on scales much smaller than $R$, corresponding to times and distances $t \ll R$, $|x| \approx R\theta \ll R$, where $\theta$ is the polar angle on the sphere. The entire difference between low and high temperature physics in effectively flat space can then be \emph{encoded completely in functions} $\rho(\lambda)$, which appear as eigenvalue densities in the more detailed description in the next two paragraphs.

At finite temperature, the integral over the gauge field can be recast into a unitary matrix model, where the projection onto singlets results from the integral over the gauge group over unitary matrices \cite{Sundborg2000} corresponding to the Polyakov loop operator, ${\cal P} \sim e^{i \int_{S^1} d\tau \alpha}$, or gauge holonomy around the thermal circle \cite{AharonyMarsanoMinwallaPapadodimasVan-Raamsdonk2004}. The distribution of the large $N$ number of matrix eigenvalues then controls the thermal behaviour.

At large $N$, the model can be solved in a saddle point approximation \cite{Sundborg2000}. This is readily achieved by introducing the eigenvalue density $\rho(\lambda)$. At low temperatures, $T < \Oo(1)$, the dominant saddle corresponds to a constant eigenvalue distribution, $\rho(\lambda) = \frac{1}{2\pi}$. This is the confined phase, with a free energy of order $N^0$. 
At intermediate temperatures whose $N$ scaling depends on the representation under consideration, there is a transition to a deconfined phase, characterised by a free energy that is extensive in $N$. At very high temperatures, the eigenvalue distribution becomes 
a delta-function\footnote{There is a subtlety here. As shown in \cite{AmadoSundborgThorlaciusWintergerst2017,AmadoSundborgThorlaciusWintergerst2018}, there are distances and times where inside correlation functions, the finite width of the eigenvalue distribution always matters. This is however unimportant for the regimes considered here.}, $\rho(\lambda) \to \delta(\lambda)$.

Correlation functions of singlet operators can be constructed through finite temperature Wick contractions. For simplicity, we focus here on the scalar singlet primary, $\Oo(t,x) = \frac{1}{N} \tr(\Phi^2(t,x))$ for scalars in the adjoint representation, whose time ordered two-point function is given by \cite{AmadoSundborgThorlaciusWintergerst2018}
\begin{align}
G(t,{\bf x}) &\equiv 
\frac{2^{d+2}\pi^{d}}{\Gamma^2(\frac{d-2}{2})} 
\left\langle \Oo(x,t) \Oo(0) \right\rangle \nonumber \\
&=\int_{-\pi}^{\pi} d\lambda\,\rho(\lambda)
\left(\sum_{m=-\infty}^{\infty}  \frac{e^{i m \lambda'}}{(\cos(t+i\beta m) 
	-  \cos\theta)^\frac{d-2}{2}}\right)^2
\,.
\label{eq:timeord}
\end{align}
where we have used rotational and time translational invariance to set one of the insertion points to zero. The pre-factor has been chosen to simplify the expression, while the operator is normalised such that its two-point function is of order $N^0$.
Eq.\eqref{eq:timeord} can formally be derived using the aforementioned Wick contraction, as well as the fact that the unitary matrix is represented in the scalar kinetic term like a temporal gauge field. 
The retarded Green's function can be extracted using its definition, $G_R(t,x) = \Theta(t) \text{Im}\, G(t,x)$. It is simple to see that the purely thermal contributions to eq. \eqref{eq:timeord} are real. An imaginary part can thus arise only from the vacuum piece, and the mixed thermal-vacuum term. Explicitly, one finds \cite{AmadoSundborgThorlaciusWintergerst2018}
\begin{multline}
G_R(t,x) = \Theta(t)\Bigg\{\text{Im}\left[\frac{1}{(\cos t -  \cos\theta)^{d-2}}\right]\\
+4\,\text{Im}\left[\frac{1}{(\cos t - \cos\theta)^{\frac{d-2}{2}}}\right]
\sum_{m=1}^{\infty} 
\text{Re}\left[ \frac{\rho_m^2}{(\cos(t+i\beta m) -  \cos\theta)^\frac{d-2}{2}}\right]\Bigg\}\,.
\label{eq:gret}
\end{multline}
where in the last line we have introduced the $k$-th Fourier cosine coefficient of the eigenvalue distribution, $\rho_k = \int \dd\lambda\, \rho(\lambda) \cos(k\lambda)$. We note that the infinite series in the second term captures all temperature dependence, and in fact is precisely that of the thermal Feynman propagator of the \emph{fundamental} scalar field, when all $\rho_m$ become equal, which is the case in the high temperature limit.

\section{Thermalisation in singlet models} \label{sec:thermalization}

A number of challenges to the OTH may be tested in thermal singlet modes in $d>2$. In this section, we describe our technical results, which support the hypothesis in even dimensions, given the adjustments we have introduced in section \ref{sec:oth}. In odd dimensions the interpretation of results is intricate, and will be deferred to the discussion \ref{sec:discussion}. The low and high temperature phases of the singlet models are qualitatively different and are discussed separately below, with equations specialised to scalars in the adjoint representation. In both cases, the concrete operator under study is the lowest dimension singlet operator $\Oo(t,x) = \frac{1}{N} \tr(\Phi^2(t,x))$.

\subsection{Low temperatures: $T < T_H$} \label{sec:lowT}
As noted in the thermal singlet model section \ref{sec:singlet models}, the eigenvalue distribution at low temperatures is constant, $\rho(\lambda) = \frac{1}{2\pi}$, and thus $\rho_0 = 1$ and $\rho_{k\neq 0} = 0$. For the retarded Green's function \eqref{eq:gret}, this implies
\begin{equation}
G_R(t,x) = \Theta(t)\,\text{Im}\left[\frac{1}{(\cos t -  \cos\theta)^{d-2}}\right]\,,
\label{eq:gretlow}
\end{equation}
in the large $N$ limit. From the explicit lack of exponentials we see that $\Oo(t,x)$ fails to thermalise at low temperatures. For completeness, let us take the ``thermodynamic limit'' of large $R$ corresponding to $t,\theta \ll 1$, and Fourier transform, thus for example obtaining
\begin{equation}\label{eq:gretlowF}
G_{R}(\omega,k) = -\log\frac{k^{2}-\omega^{2}}{\mu^{2}}\quad \mathrm{in}\, d=4,
\end{equation}
where $\mu$ is a renormalisation scale. In this Lorentz invariant expression, there is only one branch cut located at $\omega^{2}>k^{2}$ on the real line, representing a continuum of physical excitations on top of the vacuum state. That \eqref{eq:gretlowF} is analytic everywhere off the real line 
corresponds one-to-one with the fact that the corresponding expression in configuration space lacks exponentially decaying contributions. We see explicitly that it is useful to extend the notions of the non-thermalisation condition beyond poles in the frequency plane to cuts, as long as they are on the real axis. In position space, the corresponding thermodynamic limit of \eqref{eq:gretlow} involves power-law fall-off, and we will find similar fall-offs to be general consequences of free field dynamics in $d>2$ below, even for response functions of operators that display relaxation after long time in exponentially decaying terms.

The physical origin of non-thermalisation is clear from large $N$ considerations. At low $T$, one finds for the connected components of $n$-point functions
\begin{equation}
\langle \Oo(t_1,x_1) ... \Oo(t_n,x_n) \rangle_\text{conn} \sim N^{2-n}\,.
\end{equation}
In other words, $1/N$ plays the role of a coupling constant. Thus, individual $\Oo$ momentum modes are conserved to leading order in $1/N$, as long as background energy densities are not macroscopic in $N$. Composite operators like $\Oo(t,x)$ are then parametrically close to being generalised quasi-particles, and should not thermalise according to the non-thermalisation criterion. To see this, consider the effective action that generates the connected $n$-point functions of $\Oo$. Up to cubic order, and for simplicity displayed in flat space where translation invariance constrains its form, it reads
\begin{equation}
S_\text{eff} \sim \int \dd^dp \,\bigg(\Oo(p) G^{-1}(p) \Oo(-p)
+ \frac{1}{N} \int \dd^dk\,G_3^{-1}(p,k)\Oo(p)\Oo(k)\Oo(-p-k)\bigg) + ...\,,
\label{eq:effact}
\end{equation}
where $G(p) \equiv \int \dd^dx\,e^{i p x} \left\langle \Oo(x) \Oo(0) \right\rangle$ and $G_3(p,k) \equiv \int \dd^dx\dd^dy\,e^{i (px + ky)} \left\langle \Oo(x) \Oo(y) \Oo(0)\right\rangle$. 
Here, the conservation of individual momentum modes is explicit to zeroth order in $1/N$. In consequence, quasi-particles remain intact to this order. Of course, this argument is rather superficial, but can be made more precise by properly constructing the effective action, for example using collective field theory \cite{JevickiSakita1980}. 
 
By the above argument, taking into account $1/N$ corrections will reveal nontrivial features in the response functions even below the phase transition. While this requires a finite $N$ analysis, and is therefore beyond the scope of this work, even the leading order behaviour can change drastically once occupation numbers in the thermal background are of order of the inverse coupling. 
Indeed, as we will show now, this is what happens in the high temperature phase.

\subsection{High temperatures: $T \gg T_H$} \label{sec:highT}
At very high temperature, the eigenvalue distribution can be approximated by a delta-function. One thus obtains for the Fourier cosine coefficients
\begin{equation}
\label{eq:high_T_distribution}
\rho_k = 1\,.
\end{equation}
Note that one should only really expect effective thermalisation in the ``thermodynamic limit'' of large $R$, here corresponding to $t \ll 1$, $\theta \ll 1$ and $\beta \ll 1$. In this regime the retarded Green's function \eqref{eq:gret} becomes upon insertion of \eqref{eq:high_T_distribution}
\begin{multline}
G_R(t,x) = 2^{d-2}\Theta(t)\Bigg\{\text{Im}\left[\frac{1}{(x^2 - t^2)^{d-2}}\right]+4\,\text{Im}\left[\frac{1}{(x^2 - t^2)^{\frac{d-2}{2}}}\right]\\
\times
\sum_{m=1}^{\infty} 
\text{Re}\left[ \frac{1}{(x^2 - (t+i\beta m)^2)^\frac{d-2}{2}}\right]\Bigg\}\,.
\label{eq:grethigh}
\end{multline}
Clearly, the operator now responds to the thermal bath, which may induce thermalisation. 
In fact, the second term represents the cross term between vacuum and thermal propagation contributing to the response function of the quadratic composite operator $\Oo(t,x)$. 

\subsubsection{$d=4$}

To get a better understanding of the precise dynamics,  we will confine ourselves to $d = 4$, since generalisation to higher even dimensions is simple once the basic ingredients are understood. There, 
\begin{multline}
G_R(t,x) = \frac{2}{t}\Theta(t)\Bigg\{\frac{1}{2}\partial_{t}\left(\frac{1}{t}\delta(t-|x|)\right)\\
+\delta(t - |x|)
\left(\frac{2}{t^2 - x^2}+\pi\frac{\coth\frac{\pi(t+|x|)}{\beta} - \coth\frac{\pi(t-|x|)}{\beta}}{\beta |x|}\right)\Bigg\}\,.
\label{eq:grethigh2}
\end{multline}
which can be simplified to
\begin{equation}
G_R(t,x) = \frac{1}{t^2}\Bigg\{\delta'(t-|x|)
+\delta(t - |x|)\left(-\frac{2}{t}+\frac{2\pi}{\beta}\coth\frac{2\pi t}{\beta}\right)\Bigg\}\,.
\label{eq:grethigh3}
\end{equation}
Evidently, the Green's function falls off as a power law, with a power that is \emph{smaller} than in vacuum. This is in fact a manifestation of the effective dimensional reduction that is prevalent in generic thermal systems in the high temperature limit (see e.g. \cite{AppelquistPisarski1981}). However, judging by the coth term there are sub-leading exponentially decaying contributions. This may be further illuminated by Fourier transforming \eqref{eq:grethigh3}, yielding \cite{AmadoSundborgThorlaciusWintergerst2018},
\begin{equation}\label{eq:GrF}
G_{R}(\omega,k) = -2 + \frac{4\pi i}{k\beta}\log\left(\frac{\Gamma(-\tfrac{i\beta}{4\pi}(\omega-k))}{\Gamma(-\tfrac{i\beta}{4\pi}(\omega+k))}\right)-\left(\frac{2\pi i}{k\beta} - \frac{\omega}{k}\right)\log\frac{\omega+k}{\omega-k}\,.
\end{equation}
This expression allows us to map the late-time dominant behaviour of \eqref{eq:grethigh3} to the branch cut in the $\omega$ plane located between $-k$ and $k$ on the real line and the subdominant exponential decay to the branch cuts located off the real line. Similar analytic structures are discussed in \cite{HartnollKumar2007}. It can be contrasted with that of eq.\ \eqref{eq:gretlowF}.

Let us now return to how thermalisation could be consistent with the effective action arguments presented in the low temperature discussion \ref{sec:lowT}. Only large $N$ counting, which is the same at high temperature, seemed to be important. The large $N$ suppression of interactions is indeed the same as at low temperature, but the action \eqref{eq:effact} assumes the vanishing of thermal one-point functions $\langle \Oo \rangle_{\beta}$. Above the phase transition, the equilibrium background expectation value is non-zero and of order $N$, which invalidates the argument that individual $ \Oo $ momentum modes are conserved in the large $N$ limit, due to order $N^{0}$ interactions with the background. Generalised quasi-particles are then not intact in the large $N$ expansion, although their response functions are well-defined. The thermalisation of $ \Oo $ ensures that $ \Oo $ does not represent a generalised quasi-particle, by the arguments of subsection \ref{sec:non-th}. As explained above, this is consistent with large $N$ counting, thanks to the thermal condensate of $ \Oo $ above the critical temperature, which eq.\ \eqref{eq:grethigh} thus probes indirectly.

\subsubsection{General $d>2$.}

The above retarded Green's function of an operator quadratic in free fields clearly separates into a vacuum-vacuum term and a mixed vacuum-thermal term. Higher powers of free fields also decompose analogously. (Purely thermal terms will not contribute to the retarded propagator.) Now, the vacuum factors differ significantly in behaviour between odd and even dimensions. The imaginary part of $(x^2 - t^2)^{-(n+1)}$ for integer $n\geq0$ is given by
\begin{equation}
\text{Im}(x^2 - t^2)^{-(n+1)} = \frac{(\partial_{t^2})^n}{n!} \delta(t^2 - x^2)\,,
\end{equation} 
which demonstrates that the support of the retarded Green's function is confined to the light cone for even $d$, while square root branch cuts ensures support also inside the light cone for odd $d$. This is a known property of the wave equation, which evidently is inherited by thermal systems probed by composite operators built of powers of free fields. 
While the behaviour in the interior of the light cone is interesting, both for other correlation functions than the retarded Green's function, and for odd $d$, the temperature dependent term of the response function is entirely determined by the factor which multiplies a simple light cone divergence. We thus factor out its singular light cone behaviour and study the behaviour of what might be called the position space ``residue'' of the singularity, by abuse of terminology. 

The light-cone factor isolated from eq.\ \eqref{eq:grethigh} is then
\begin{equation}\label{eq:lc factor}
G_R(t,x)^{{\text{lc}}} = 2^{d-2}\Theta(t) \sum_{m=1}^{\infty} 
\text{Re}\left[ \frac{1}{(x^2 - (t+i\beta m)^2)^\frac{d-2}{2}}\right]\equiv 2^{d-2}\Theta(t) \tilde{S}_{d}(t,x)\,.
\end{equation}
This expression lends itself to a comparatively uniform treatment independently of dimension, and it measures the effect of the heat bath on the light cone in position space. The functions $\tilde{S}_{d}(t,x)$ and its light cone limit $S_{d}(t)$ are discussed in the appendix \ref{sec:lcapp}.

In even dimensions the calculation confirms thermalisation on the light cone, essentially by expressing $S_{d}(t)$ as an expansion in modified Bessel functions
\begin{equation}\label{eq:bessel series}
 \sum_{m \in \mathbb{Z}\setminus\{0\}} \frac{2e^{-2m\pi\bar{t}}\pi^{\frac{d-3}{2}}|m|^{\frac{d-3}{2}}K_{\frac{d-3}{2}}\left(2\pi |m| t/\beta \right)}{\left(t/ \beta\right)^{\frac{d-3}{2}}},
\end{equation}
each of which equals a decaying exponential times a terminating sum of inverse powers for even dimensions (i.e.\ half-integer orders of the Bessel function). Details concerning finiteness of the expressions are also given in the appendix \ref{sec:lcapp}. For all even dimensions the positive $m$ terms in the series above explicitly yield exponentially decaying terms, which are of a form that cannot cancel with other exponentials.  The negative $m$ terms produce power law fall-off. Thus, the response functions signal partial thermalisation on the light cone in even dimensions. 

The odd-dimensional case is significantly more subtle.
A similar treatment of the Bessel function terms in series \eqref{eq:bessel series} now leads to an asymptotic expansion in inverse powers for each $m$, which does not terminate. Hence, it is far from clear what significance to attach to exponentially small terms. If the asymptotic series is truncated, the error terms will be larger than the exponentials we have extracted, even if exponentially improved expansions are used \cite{NIST:DLMF}.

\section{Discussion} \label{sec:discussion}

Our description of an important class of non-thermalising operators as generalised free field operators in section \ref{sec:thermalization} connects to the intuition that such operators should not thermalise. Generally, however, naive intuition is treacherous, and our study is founded on the observation that free field equations of motion do not generally guarantee absence of relaxation for operators which are non-linear in free fields. Composite operators are regularised operators belonging to this class, and as described in the introduction \ref{sec:intro}, they have in many instances been shown to display the decay which we take to define thermalisation. The idea of the OTH is that the implication could go in the other direction: Operators that do not thermalise partially would have to be generalised quasi-particle operators. Or equivalently, any other operators  thermalise partially.

Known thermal behaviour of singlet models motivated a closer study of response functions in order to compare with the OTH in dimensions $d>2$. The $d > 2$ treatment of \cite{Sabella-GarnierSchalmVakhtelZaanen2019} is somewhat less detailed than the $d=2$ discussion, and we were able  to resolve new even/odd dimension differences in eqs.\ (\ref{eq:lc factor}-\ref{eq:bessel series}) from the high temperature phase of the singlet model response functions. To get expressions which depend analytically on $d$, it proved important to focus on the light cone. Thereby, the qualitative difference between the support for the Green's functions in the light cone, related to the absence or not of Huygens' principle were factored out.

In the large $N$ limit, where there is a phase transition, and below the critical temperature, the response function \eqref{eq:gretlow} lacks exponentially decaying terms and the individual momentum modes are independently conserved. This means non-thermalisation and also the presence of generalised quasi-particles of definite momenta, as expected from the general non-thermalisation results. Clearly, we can only expect a precise match to the general operator thermalisation theory, described in section \ref{sec:non-th}, to be valid at leading order in small $1/N$. To the extent that the generalised quasi-particle picture holds, we can rely on non-thermalisation. Indeed, the idea that the general theory applies parametrically close to ideal cases makes the results much more powerful. In this example, we see how it works.

Above the critical temperature, the response functions develop exponentially decaying terms, as for example in the $d=4$ expressions \eqref{eq:grethigh3} and \eqref{eq:cosh}. These examples clearly show how power law tails and damped exponentials combine non-trivially. Indeed, such terms which generally appear in $d>2$ motivate us to consider \emph{partial} operator thermalisation. This partial thermalisation concept also simplifies the non-thermalisation results for stable thermal quasi-particles, by allowing branch cuts in the thermodynamic limit as in the paragraph after eq.\ (\ref{eq:matrix elements}).

To conclude the match of the OTH and singlet model response functions we should now argue that the operators we consider fail to be generalised quasi-particle operators above the critical temperature. Without going deeply into the physics of singlet models, we have found a suitable mechanism, namely that the background condensate in the high temperature phase modifies the propagation of perturbations at order $N^{0}$ which is too much for a generalised free field, unless there is extreme fine-tuning.

Partial thermalisation diffuses the dichotomy between thermalising and non-thermalis\-ing  
operators to some degree, but in even dimensions calculations like \eqref{eq:besselsd} and \eqref{eq:s4calc} demonstrate the general structure from modified Bessel functions of order $\frac{d-3}{2}$. At half integer order the resulting functions are simple polynomials of exponentials $\exp{(-4\pi t/\beta)}$ and  powers of $\beta/t$. The further sums in the thermal response functions primarily gives rise to an infinite series of higher order terms $\exp{(-4\pi n t/\beta)}$, where $n$ are integers, but the expansion in $\beta/t$ terminates and the damped exponential terms can be distinguished from the resulting polynomial. Thermalisation can be confirmed although with a bit more work than if power law tails had not been present. This comparatively simple procedure works in even dimensions, when the whole effect of the induced thermalisation is confined to the light cone by Huygens' principle.

In odd dimensions Huygens' principle does not apply and some of the induced thermalisation diffuses into the interior of the light cone. The series encoding the thermalisation on the light cone, which corresponds to the polynomial in $\beta/t$, now fails to terminate. Instead it produces an infinite asymptotic expansion controlled by the asymptotic expansion of modified Bessel functions of integer order. The resulting series in $\beta/t$ is divergent and the task to identify sub-leading exponentials becomes quite subtle. The potential meaning of exponential terms can only be ascertained within a larger framework, such as the study of resurgence of asymptotic series. In such a framework one should be able to assign a meaning to partial thermalisation of composite operators in odd-dimensional free field theories, but a firm conclusion is beyond the scope of the present work. 

Tentatively, the Borel summability of modified Bessel asymptotic expansion indicates that there are no exponential correction terms in its asymptotic expansions, which would suggest that the exponential terms we actually find in the appendix are not masked, but on the other hand error terms of even doubly improved asymptotic series are of the same order as the sub-leading exponential terms.

\section{Conclusions} \label{sec:conclusions}

We have refined the operator thermalisation concept and the OTH, and related it to generalised free fields. Except in the special case $d=2$, operator thermalisation is generally incomplete and partial, since there are power law tails that dominate exponentially decaying terms at late times. This finding establishes the intermediate nature of thermalisation in free field theories: while exponential relaxation is ubiquitous, it typically coexists with the more unyielding time dependence expected from the presence of conservation laws. 

In our model system, large $N$ singlet models, we have found both non-thermalising and thermalising behaviour of the same operator: generalised quasi-particle behaviour without exponential relaxation below the critical temperature, and thermalising exponential behaviour above the critical temperature. Importantly, the operator thermalisation concepts turn out to be applicable to operators which only satisfy the theoretical conditions in a limit, in this case when $1/N$ vanishes. This enhances the scope of our analysis.

The analysis is comparatively straightforward in even dimensions, where Huygens' principle holds and ensures that the thermalised responses induced by a heat bath are localised to the light cone. In contrast, the thermalised responses in odd dimensions are quite intricate due to their distribution over the forward light cone and the whole of its interior. We refrain from formulating a definite conclusion in odd dimensions, since we believe in a deeper conceptual analysis. The importance of simultaneous infinite expansions in inverse powers, and decaying exponentials, of time, suggests resurgent analysis. A connection between thermalisation of integrable systems and resurgence may find further applications.

Some properties of singlet models that are highlighted by our study generate further questions. For example, an efficient description of the high temperature phase remains elusive. We expect that all standard composite operators will thermalise and no longer represent generalised quasi-particles. The fundamental free fields $\Phi$ describe the thermodynamics of the high temperature ``deconfined'' limit well, but they do not represent physical singlet states. Do they provide the best description, or are there better alternatives? There are also holographic gravity duals to these questions, since singlet models are limits of large N gauge theories, some of which are conformal.

Finally, we find it inspiring to contemplate other conformal or integrable systems, in particular in odd dimensions, where resurgence appears to be fundamental.

\acknowledgments{
It is a pleasure to thank P. Sabella-Garnier, K. Schalm and J. Zaanen for discussions.
The work of BS was supported by the Swedish Research Council contract
DNR-2018-03803 and that of NW by FNU grant number DFF-6108-00340.
}

\appendix

\section{Thermal contribution to light cone $G_{R}$ in any $d$} \label{sec:lcapp}
From equation (\ref{eq:grethigh}), the temperature-dependent part of the retarded Green's function is determined by the sum
\begin{equation}\label{eq:sum}
\tilde{S}_{d}(t,x) = \sum_{m=1}^{\infty}\text{Re}\left[ \frac{1}{(x^2 - (t+i\beta m)^2)^\frac{d-2}{2}}\right]\,.
\end{equation}
For later comparison, in $d=4$ we have
\begin{align}
\tilde{S}_{4}(t,x) &= \sum_{m=1}^{\infty} \text{Re}\left[ \frac{1}{x^2 - (t+i\beta m)^2}\right] = \frac{1}{2}\left(\sum_{m=-\infty}^{\infty}\left[ \frac{1}{x^2 - (t+i\beta m)^2}\right] -\frac{1}{x^{2}-t^{2}}\right) \notag\\
&= \frac{1}{2}\left(\frac{\pi}{2\beta x}\left[\coth\frac{\pi(t+x)}{\beta} - \coth\frac{\pi(t-x)}{\beta}\right]-\frac{1}{x^{2}-t^{2}}\right)\,.
\end{align}
Since in even $d$ the retarded Green's function only has support on the light cone we will only evaluate the sum there. We have
\begin{equation}\label{eq:cosh}
S_{4}(t) =  \frac{\pi}{4\beta t}\coth\frac{2\pi t}{\beta}-\frac{1}{8 t^{2}}\,.
\end{equation}

The sum \eqref{eq:sum} can be rewritten using the formula
\begin{equation}
\frac{1}{y^{\alpha}} = \frac{1}{\Gamma(\alpha)}\int_{0}^{\infty}s^{\alpha-1}e^{-ys} \dd s\,.
\end{equation}
We thus obtain
\begin{align}
\tilde{S}_{d}(t,x) &= \sum_{m=1}^{\infty}\text{Re}\left[ \frac{1}{(x^2 - (t+i\beta m)^2)^\frac{d-2}{2}}\right] \notag \\
&= \sum_{m=1}^{\infty}\text{Re}\left[\frac{1}{\Gamma(\frac{d-2}{2})}\int_{0}^{\infty}s^{\frac{d-4}{2}}e^{-(x^2 - (t+i\beta m)^2)s} \dd s\right]\notag\\
&= \frac{1}{\Gamma(\frac{d-2}{2})}\sum_{m=1}^{\infty}\text{Re}\left[\int_{0}^{\infty}s^{\frac{d-4}{2}}e^{-(x^2 - t^{2} + \beta^{2} m^{2} -2i t \beta m)s} \dd s\right]\,.
\end{align}
Restricting ourselves to the light cone we obtain 
\begin{align}\label{eq:int}
S_{d}(t) &= \frac{1}{\Gamma(\frac{d-2}{2})}\int_{0}^{\infty}s^{\frac{d-4}{2}}\left(\sum_{m=1}^{\infty}e^{- \beta^{2} m^{2} s} \cos(2t \beta sm)\right)\,\dd s\notag\\
&= \frac{1}{2\Gamma(\frac{d-2}{2})}\int_{0}^{\infty}s^{\frac{d-4}{2}}(\vartheta(t\beta s, e^{-s\beta^{2}})-1)\,\dd s\notag\\
&= \frac{1}{2\beta^{d-2}\Gamma(\frac{d-2}{2})}\int_{0}^{\infty}r^{\frac{d-4}{2}}(\vartheta(r\bar{t}, e^{-r})-1)\,\dd r\notag\\
\end{align}
where $\vartheta(z,q)$ is the third Jacobi theta function and $\bar{t} = t/\beta$.
Using the modular transformation property of $\vartheta$, we obtain
\begin{align}\label{eq:modtrans}
\vartheta(r\bar{t},e^{-r}) &= \sqrt{\frac{\pi}{r}}e^{-r\bar{t}^{2}}\vartheta(i\pi\bar{t},e^{-\frac{\pi^{2}}{r}}) = \sqrt{\frac{\pi}{r}}e^{-r\bar{t}^{2}}\left(1+ 2\sum_{m=1}^{\infty}e^{-\frac{\pi^{2}m^{2}}{r}}\cosh(2\pi m\bar{t})\right)\notag\\
&=\sqrt{\frac{\pi}{r}}\sum_{m=-\infty}^{\infty}e^{-r(\bar{t}+\frac{\pi m}{r})^{2}}\,.
\end{align}
Substituting for this in the integral \eqref{eq:int} yields
\begin{align}
S_{d}(t) &= \frac{\sqrt{\pi}}{2\beta^{d-2}\Gamma(\frac{d-2}{2})}\int_{0}^{\infty}r^{\frac{d-5}{2}}\sum_{m=-\infty}^{\infty}e^{-r(\bar{t}+\frac{\pi m}{r})^{2}}\,\dd r \notag\\
&- \frac{1}{2\beta^{d-2}\Gamma(\frac{d-2}{2})}\int_{0}^{\infty}r^{\frac{d-4}{2}}\,\dd r\,.
\end{align}
The last term is divergent but will be canceled by a divergence stemming from the sum. In order to allow for such a cancellation we regularise the integral by introducing an exponential suppression $e^{-r/R}$ with $R$ taken to infinity after performing the integral. We have
\begin{align} \label{eq:besselsd}
S_{d}(t) &= \lim_{R\rightarrow\infty}\Bigg[\frac{\sqrt{\pi}}{2\beta^{d-2}\Gamma(\frac{d-2}{2})}\sum_{m=-\infty}^{\infty}\int_{0}^{\infty}r^{\frac{d-5}{2}}e^{-r(\bar{t}+\frac{\pi m}{r})^{2}-\frac{r}{R}}\,\dd r \notag\\
&- \frac{1}{2\beta^{d-2}\Gamma(\frac{d-2}{2})}\int_{0}^{\infty}r^{\frac{d-4}{2}}e^{-\frac{r}{R}}\,\dd r\Bigg]\notag\\
&= \lim_{R\rightarrow\infty}\Bigg[\frac{\sqrt{\pi}}{2\beta^{d-2}\Gamma(\frac{d-2}{2})}\left(\frac{\Gamma(\frac{d-3}{2})}{\bar{t}^{d-3}} + \sum_{m \in \mathbb{Z}\setminus\{0\}} \frac{2e^{-2m\pi\bar{t}}\pi^{\frac{d-3}{2}}|m|^{\frac{d-3}{2}}K_{\frac{d-3}{2}}\left(2\pi |m| \sqrt{\bar{t}^{2}+\frac{1}{R}}\right)}{\left(\bar{t}^{2}+\frac{1}{R}\right)^{\frac{d-3}{4}}}\right)\notag\\
&- \frac{R^{\frac{d-2}{2}}}{2\beta^{d-2}}\Bigg]\,,
\end{align}
which is valid for $d>3$. As we are interested in large $t$-behaviour we employ the asymptotic expansion
\begin{align}
K_{\alpha}(z) &= \sqrt{\frac{\pi}{2z}}e^{-z}\left(1+\frac{4\alpha^{2}-1}{8z} + \frac{(4\alpha^{2}-1)(4\alpha^{2}-9)}{2!(8z)^{2}}\right.\notag\\
&+ \left.\frac{(4\alpha^{2}-1)(4\alpha^{2}-9)(4\alpha^{2}-25)}{3!(8z)^{3}}+\ldots\right)\equiv\sqrt{\frac{\pi}{2z}}e^{-z}\sum_{n=0}^{\infty}\frac{a_{n}(\alpha)}{z^{n}}\,.
\end{align}
We note that for half-integer $\alpha$ the series terminates at a finite number of terms. This corresponds to even $d$. Substituting for this in \eqref{eq:besselsd} yields
\begin{align}\label{eq:asymp}
S_{d}(t) 
&= \lim_{R\rightarrow\infty}\Bigg[\frac{\sqrt{\pi}}{2\beta^{d-2}\Gamma(\frac{d-2}{2})}\Bigg(\frac{\Gamma(\frac{d-3}{2})}{\bar{t}^{d-3}} + \frac{\pi^{\frac{d-3}{2}}}{\bar{t}^{\frac{d-2}{2}}}\sum_{n=0}^{\infty}\frac{a_{n}(\frac{d-3}{2})}{(2\pi \bar{t})^{n}}\sum_{m =1}^{\infty} m^{\frac{d-4}{2}-n}e^{-4\pi m\bar{t}}\notag\\
&+ \pi^{\frac{d-3}{2}}\sum_{n=0}^{\infty}\frac{a_{n}(\frac{d-3}{2})}{(2\pi)^{n} \left(\bar{t}^{2}+\frac{1}{R}\right)^{\frac{d+2n-2}{4}}}\sum_{m =1}^{\infty} m^{\frac{d-4}{2}-n}e^{-2\pi m\left(\sqrt{\bar{t}^{2}+\frac{1}{R}}-\bar{t}\right)}\Bigg)- \frac{R^{\frac{d-2}{2}}}{2\beta^{d-2}}\Bigg]\,.
\end{align}
As a check we set $d=4$ to compare with \eqref{eq:cosh}. We obtain
\begin{align} \label{eq:s4calc}
S_{4}(t) &= \lim_{R\rightarrow\infty}\Bigg[\frac{\pi}{2\beta^{2}}\left(\frac{1}{\bar{t}} + \sum_{m \in \mathbb{Z}\setminus\{0\}} \frac{e^{-2\pi\left(m\bar{t}+|m|\sqrt{\bar{t}^{2}+\frac{1}{R}}\right)}}{\sqrt{\bar{t}^{2}+\frac{1}{R}}}\right)- \frac{R}{2\beta^{2}}\Bigg]\notag\\
&= \lim_{R\rightarrow\infty}\Bigg[\frac{\pi}{2\beta^{2}}\left(\frac{1}{\bar{t}} + \frac{1}{\bar{t}}\sum_{m = 1}^{\infty} e^{-4\pi m\bar{t}}+\frac{1}{\sqrt{\bar{t}^{2}+\frac{1}{R}}}\sum_{m = 1}^{\infty} e^{-2\pi m\left(\sqrt{\bar{t}^{2}+\frac{1}{R}}-\bar{t}\right)}\right)- \frac{R}{2\beta^{2}}\Bigg]\notag\\
&= \lim_{R\rightarrow\infty}\Bigg[\frac{\pi}{2\beta^{2}}\left(\frac{1}{\bar{t}} + \frac{1}{\bar{t}}\sum_{m = 1}^{\infty} e^{-4\pi m\bar{t}}+\frac{R}{\pi}-\frac{1}{4\pi \bar{t}^{2}}-\frac{1}{2\bar{t}}\right)- \frac{R}{2\beta^{2}}\Bigg]\notag\\
&= \frac{\pi}{2\beta^{2}}\left(\frac{1}{2\bar{t}} + \frac{1}{\bar{t}}\sum_{m = 1}^{\infty} e^{-4\pi m\bar{t}}-\frac{1}{4\pi \bar{t}^{2}}\right)\notag\\
&=  \frac{\pi}{4\beta t}\coth\frac{2\pi t}{\beta}-\frac{1}{8 t^{2}}\,,
\end{align}
which agrees with \eqref{eq:cosh}.

For odd dimensions we investigate \eqref{eq:asymp} for $d=3$ even though the expression is technically only valid for $d>3$ because of the logarithmic divergence in the first term. 
\begin{align}
S_{3}(t) &= \frac{1}{2\beta}\Bigg(\Gamma(0) + \frac{1}{\sqrt{\bar{t}}}\sum_{n=0}^{\infty}\frac{a_{n}(0)}{(2\pi \bar{t})^{n}}\sum_{m =1}^{\infty} \frac{e^{-4\pi m\bar{t}}}{m^{n+\frac{1}{2}}} + \frac{1}{\sqrt{\bar{t}}}\sum_{n=0}^{\infty}\frac{a_{n}(0)\zeta(n+\frac{1}{2})}{(2\pi \bar{t})^{n}}\Bigg)\,.
\end{align}

As can be seen, the asymptotic power series does not terminate and the sub-leading exponentials are masked, suggesting that more powerful asymptotic or resurgence method should be employed. 

\bibliographystyle{utphys}
\bibliography{OTHBibliography}

\end{document}